\documentclass[twocolumn]{aastex631}

\usepackage{placeins}
\usepackage{xspace}
\usepackage{natbib}
\bibpunct{(}{)}{;}{a}{}{,}
\usepackage{graphicx}
\usepackage{xcolor}
\usepackage{multirow}
\usepackage{amsmath}
\usepackage{lipsum}

\newcommand{\NH}{\ensuremath{N_{\mathrm{H}}}\xspace}
\newcommand{\nH}{\ensuremath{n_{\mathrm{H}}}\xspace}

\newcommand{\amol}{{\tt amol}\xspace}
\newcommand{\hot}{{\tt hot}\xspace}
\newcommand{\pow}{{\tt pow}\xspace}

\newcommand{\pion}{{\tt pion}\xspace}
\newcommand{\dbb}{{\tt dbb}\xspace}
\newcommand{\comt}{{\tt comt}\xspace}
\newcommand{\refl}{{\tt refl}\xspace}
\newcommand{\etau}{{\tt etau}\xspace}

\newcommand{\sectors}{{\tt sectors}\xspace}
\newcommand{\xmm}{{XMM-\textit{Newton}}\xspace}
\newcommand{\chandra}{\textit{Chandra}\xspace}
\newcommand{\spex}{{\tt SPEX}\xspace}

\newcommand{\feiii}{\ion{Fe}{3}\xspace}
\newcommand{\feiv}{\ion{Fe}{4}\xspace}

\newcommand{\oii}{\ion{O}{2}\xspace}
\newcommand{\oiii}{\ion{O}{3}\xspace}
\newcommand{\oiv}{\ion{O}{4}\xspace}

\newcommand{\ovii}{\ion{O}{7}\xspace}
\newcommand{\oviii}{\ion{O}{8}\xspace}

\newcommand{\neiii}{\ion{Ne}{3}\xspace}
\newcommand{\neiv}{\ion{Ne}{4}\xspace}
\newcommand{\neix}{\ion{Ne}{9}\xspace}
\newcommand{\nex}{\ion{Ne}{10}\xspace}

\newcommand{\sxiv}{\ion{S}{14}\xspace}
\newcommand{\sxv}{\ion{S}{15}\xspace}
\newcommand{\fexix}{\ion{Fe}{19}\xspace}
\newcommand{\fexx}{\ion{Fe}{20}\xspace}
\newcommand{\fexxi}{\ion{Fe}{21}\xspace}

\begin{document}

% -------------- Affiliation and title ------------ %
\title{Connecting dust and outflows in AGN: the intriguing case of NGC 6860}

\author[0000-0002-1049-3182]{I. Psaradaki}
\affiliation{MIT Kavli Institute for Astrophysics and Space Research, 70 Vassar Street, Cambridge, MA 02139}
\affiliation{SRON Netherlands Institute for Space Research, Niels Bohrweg 4, 2333 CA Leiden, the Netherlands}
\author{M. Mehdipour}
\affiliation{Space Telescope Science Institute, 3700 San Martin Dr, Baltimore, MD 21218, USA}
\author{D. Rogantini}
\affiliation{Department of Astronomy and Astrophysics, The University of Chicago, Chicago, IL 60637}
\author{E. Costantini}
\affiliation{SRON Netherlands Institute for Space Research, Niels Bohrweg 4, 2333 CA Leiden, the Netherlands} 
\affiliation{Anton Pannekoek Institute, University of Amsterdam, Postbus 94249, 1090 GE Amsterdam, The Netherlands}
\author{N. Schulz}
\affiliation{MIT Kavli Institute for Astrophysics and Space Research, 70 Vassar Street, Cambridge, MA 02139}
\author{S. Zeegers}
\affiliation{European Space Agency (ESA), European Space Research and Technology Centre (ESTEC), Keplerlaan 1, 2201 AZ Noordwijk, The Netherlands}
\author{E. Caruso}
\affiliation{SRON Netherlands Institute for Space Research, Niels Bohrweg 4, 2333 CA Leiden, the Netherlands}
\affiliation{Anton Pannekoek Institute, University of Amsterdam, Postbus 94249, 1090 GE Amsterdam, The Netherlands}

\begin{abstract}
Cosmic dust plays a crucial role in the evolution of galaxies, significantly influencing star formation and the interstellar medium. However, in active galactic nuclei (AGN), the role and origin of dust remain poorly understood. High-resolution X-ray spectroscopy is a powerful tool for probing the properties of dust in AGN.
NGC 6860, an X-ray bright type-1 quasar, is an ideal target for investigating the connection between dust and winds in AGN. It exhibits reddening and X-ray absorption by both dust and winds. By modeling high-resolution X-ray spectra from \xmm and \chandra observations, we determine the properties of dust and outflows in this AGN.
Our analysis finds four photoionized components, outflowing with velocities of 50--300 km~s$^{-1}$. The first two are relatively highly ionized with $\log\xi=3.4$ and $\log\xi=2.9$. The results of our photoionization modeling suggest that these two components are thermally unstable. The third component is ionized, with $\log\xi=2.3$ and is located further away from the central black hole. The fourth component is less ionized, and is possibly located in the host galaxy. The application of dust models enables us to probe the abundance and location of the dust in NGC 6860. Our findings suggest that dust absorption and reddening originates from the extended narrow-line region and its host galaxy.

\end{abstract}

\section{Introduction} \label{sec:intro}

% Write about the dust
Cosmic dust grains are small solid particles, typically ranging in size from a few nanometers to a few micrometers, composed mainly of heavy elements. They are widespread in the universe and can be used to trace the evolutionary paths of planets, stars or even black holes. Dust grains cool down interstellar clouds through radiative transfer,
enabling star formation, provide the surface for chemical reactions, and are the building
blocks from which planetesimals are formed in planet-forming disks around young stars. Dust, mainly characterised by its chemical composition, size and lattice structure (crystallinity), contains much of the heavy elements, such as Fe, O, Si, Mg, and C, in solid form (\citealt{Drainebook}).

% Write about XAFS
In the last decades, high-resolution X-ray absorption spectroscopy has proven to be a powerful tool for studying the cosmic dust (e.g. \citealt{Zeegers2019}, \cite{Rogantini2020}, \citealt{Psaradaki2022}, \citealt{Costantini2022}, \citealt{Corrales2024}).
In the proximity to the photoelectric absorption edges, one can observe the X-ray Absorption Fine Structures (XAFS), which are oscillatory modulations and appear when an X-ray photon interacts with a dust grain. This interaction modulates the wave-function of the photo-electron due to constructive and destructive interferences \citep{Newville}. The XAFS's shape contains information about the dust grain chemical composition, size and lattice structure. X-ray absorption spectroscopy of interstellar dust and gas, is nowadays a very powerful way to study the chemical enrichment of galaxies. 

% Write about Dust in AGN
The origin and properties of winds in active galactic nuclei (AGN) are still uncertain. Dust plays a crucial role in their dynamics and observational characteristics. In AGN, especially Seyfert 2 galaxies, dust contributes to the obscuration and reddening of the AGN light. The dust torus can block the direct view of the central black hole, absorbing ultraviolet and optical radiation from the accretion disk around the supermassive black hole. This absorbed energy is then re-emitted in the infrared, which significantly impacts the AGN’s spectral energy distribution. The optically thick dusty torus plays a major role in determining how we view the AGN, depending on our viewing angle relative to the orientation of this obscuring structure. (e.g., \citealt{Komossa1997}; \citealt{Reynolds1997}; \citealt{Lee2001}; \citealt{Crenshaw2001}).

Dust in AGN can also play a critical role in the chemical enrichment of the host galaxy's interstellar medium (e.g., \citealt{Oppenheimer2006}). Observed relationships between supermassive black holes (SMBH) and their host galaxies suggest that these two systems co-evolve through feedback mechanisms. AGN-driven winds are crucial to this co-evolution \cite[e.g. ][]{diMatteo2005}. However, significant gaps remain in our understanding of AGN outflows, leading to uncertainties about their exact role and impact on galaxy evolution.

The origins, structure, and mechanisms that drive AGN winds remain largely unclear. Depending on their ionizations states, different outflows can have different origin and acceleration mechanisms (\citealt{Crenshaw2003, Laha2020}). Some of the proposed mechanisms include thermally-driven winds (e.g. \citealt{Begelman1983}, \citealt{Krolik2001}, \citealt{Proga2002}), radiatively-driven (\citealt{Dorodnitsyn2008}), or magnetically-driven processes (e.g. \citealt{Kazanas2012}, \citealt{Fukumura2015}, \citealt{Fukumura2018}).
Understanding the role of dust in these processes is significant because it influences the dynamics and cooling of interstellar gas, contributes to star formation, and affects the overall chemical evolution of galaxies.

Warm-absorbers, commonly detected in the UV and X-ray spectra of bright AGN, have been found to be consistent with being torus winds (e.g., \citealt{Kaastra2012}; \citealt{Mehdipour2018}), suggesting that dust could be mixed with these outflows. Notably, infrared radiation pressure on dust grains can enhance these torus winds (\citealt{Dorodnitsyn2011}). Therefore, establishing the existence of dust in these systems is crucial for understanding the driving mechanisms of the outflows and defining their environmental impact. This knowledge is essential for understanding the contribution of winds to AGN feedback and the co-evolution of supermassive black holes and their host galaxies.

Observational evidence indicates that AGN winds play a significant role in transporting dust into the interstellar medium (ISM). For example, in the case of ESO 113-G010 \citep{Mehdipour2012}, the presence of dust within the AGN wind was inferred from changes in the AGN emission, from the dust-to-gas ratio, and the characteristics of the wind. Additionally, lower ionization states of warm absorbers are believed to harbor dust \cite[][and references therein]{Laha2020}, which is crucial for understanding the feedback momentum and kinematics driven by the central black hole \citep{Fabian2008, Ishibashi2018}. Further supporting this, \citet{MehdipourCostantini2018} performed a multi-wavelength study of the reddened quasar IC 4329A and identified a dust component linked to outflows from the AGN torus.

We undertake a case study for NGC 6860 to study the nature of the dust in the vicinity of the AGN. Dust-reddened AGN are often too faint for X-ray spectroscopy due to strong absorption, making it challenging to find a feasible target. For our X-ray spectroscopic investigation of dust in winds, the AGN target must satisfy several criteria: it must be sufficiently bright in X-rays, there is presence of ionised gas, show evidence of dust features in both X-rays and infrared, demonstrate the presence of an AGN wind, and have minimal reddening or absorption contamination from the Milky Way along our line of sight. NGC 6860 is one of the most suitable AGN that meets all these requirements.

The total Galactic \NH in the line of sight is $\rm 3.58 \times 10^{20} cm^{-2}$ \citep{Willingale2013}. NGC 6860 exhibits infrared spectral features indicative of dust, as observed in Spitzer/IRS data (\citealt{Wu2009}, \citealt{Gallimore2010}). The data revealed 6.2 and 11.2 $\mu$m dust features linked to carbon within the galaxy. The reddening in the host galaxy is found to be $E(B-V)=0.72$ \citep{Lipari1993}, which is significantly higher than the Milky Way reddening in our line of sight: E(B-V)=0.037. \citet{Lipari1993} also concluded from the optical, near-IR and far-IR data that in the nuclear regions of NGC 6860, the ionising source has a composite nature, a variable Seyfert 1 nucleus embedded in an intense and dusty star formation environment. This indicates significant dust extinction in NGC 6860.

The paper is organized as follows: In Section \ref{data_red}, we describe the reduction process of the \xmm and \chandra data. In Section \ref{fit}, we present the broadband spectral fitting techniques used to obtain the Spectral Energy Distribution (SED) and model the ionised outflows (or 'warm absorbers') and dust absorption. Section \ref{discussion} discusses our results, and Section \ref{conclusions} presents our conclusions.

\section{ Observational data and reduction} \label{data_red}

We use high-resolution X-ray spectroscopic data from the Reflection Grating Spectrometer \cite[RGS,][]{denHerder} on board XMM-Newton and Chandra data taken with the High Energy Transmission Grating Spectrometer \cite[HETGS,][]{hetgs}. We further use Optical Monitor data (OM) from XMM-Newton in order to constrain the SED.

All data used in this study are listed in Table \ref{observation_log}. 
We processed the XMM-Newton data using the Science Analysis System (SAS v20.0). The RGS was used in the standard Spectro+Q mode in our observation. Data processing included executing the {\tt rgsproc} pipeline task, extracting both source and background spectra, and generating response matrices. Time intervals with background count rates exceeding 0.1 count/s in CCD number 9 were filtered out. The first-order RGS1 and RGS2 spectra are simultaneously fitted over the 6--37 Å band.

% ------------------------ OBS LOG ------------------------- %
\begin{table*}
\caption{Observation log. } 
\label{observation_log}   
\centering
\scalebox{1}{
\begin{tabular}{c c c c c }     
\hline   
\hline
Satellite                       & obs ID         &     instrument/mode   &   exp. time (ks) & date of obs.   \\
\hline   
\hline                
{\xmm}         &   0903030101   &   RGS \& OM              &    120  & 18/03/2023          \\
\hline                
{\multirow{7}{*}{\chandra}}    &     22620        &    HETGS/TE          &    25     & 01/12/2020  \\
 		 &      22621            &      HETGS/TE    &     17  & 09/12/2019         \\
     	   &      22969            &      HETGS/TE    &       30  & 07/07/2021       \\
 		 &       23095           &      HETGS/TE       &      33  & 13/12/2019        \\
 		 &        24779          &      HETGS/TE       &        30  & 27/02/2021      \\
 		 &        24780          &      HETGS/TE        &         30  & 29/11/2020      \\
 		 &        24877          &      HETGS/TE       &            25  & 01/12/2020   \\
\hline                
\end{tabular}}
\begin{flushleft}
\footnotesize{Note: TE is an acronym for Timed Exposure mode. The XMM-Newton OM data were taken with the primary photometric filters $V$, $B$, $U$, $UVW1$, $UVM2$, and $UVW2$.  }
\end{flushleft}
\end{table*}

The XMM-Newton OM data, operating in the Science User Defined mode, were taken with the primary photometric filters $V$, $B$, $U$, $UVW1$, $UVM2$, and $UVW2$. We applied a circular aperture with a diameter of 12'' for our photometry, which is the optimal size based on the OM calibration.

The Chandra HETG data were reduced using the Chandra Interactive Analysis of Observations \cite[CIAO,][]{Fruscione2006} v4.15 software and Calibration Database (CALDB) v4.10.7. The {\tt chandra\_repro} script of CIAO and its associated tools were used for the reduction of the data and production of the final grating products (PHA2 spectra, RMF and ARF response matrices). To improve the signal-to-noise ratio and enhance our ability to detect absorption lines, we need to combine the Chandra observations. The data are divided into numerous short exposures, but after analyzing the overall flux variations between them, we find no significant evidence of variability. Additionally, the spectral shape of the continuum, as determined from our HETG fit, aligns well with the lower-energy shape observed with RGS. This suggests that NGC 6860 does not experience substantial variability during our observation period that would impact our modeling.

\section{Spectral fitting}
 \label{fit}

  For our modelling we use the software SPEctral X-ray and UV modelling and analysis, \spex, version 3.06.01\footnote{\url{http://doi.org/10.5281/zenodo.2419563}} (\citealt{Kaastra1996, Kaastra2018}). To evaluate the
goodness of our fit we adopt $C$-statistics which is denoted as
$C_{\rm stat}$ in the rest of the manuscript. SPEX uses $C_{\rm stat}$ as described
in \citet{Cash1979}, but with an addition which allows an estimate of
the goodness of the fit (for details see \citealt{Kaastra2017}). The errors are provided at 1$\sigma$ (68\%) confidence level. Also, in our analysis we use proto-Solar abundance units from \citet{Lodders2009}. 

\subsection{Optical-UV continuum and reddening} 
\label{sect_optical}

The optical/UV continuum was modeled using a standard disk blackbody component ({\dbb} in \spex). This models adopts a geometrically thin, optically thick, Shakura-Sunyaev accretion disk model \citep{Sha73}. We fix the temperature of the {\tt dbb} model to 15 eV and fit the normalization of this component. This continuum model is also reddened. We use the {\tt ebv} model in \spex which takes into account internal reddening in the AGN. Another {\tt ebv} component is used for the Milky Way reddening. For the internal reddening, we used the Small Magellanic Cloud (SMC) extinction curve of \citet{Gord03}, which is called `SMC Bar Average'. Such SMC extinction curve is commonly thought to describe dust extinction in AGN \cite[e.g.][]{Hopkins2004, Mehdipour2021}. The $E(B-V)$ parameter of this component is fitted to the OM data. For the Milky Way reddening, we use the extinction curve of \citet{Car89}, including the update for near-UV given by \citet{ODo94}. The ${E(B-V)}$ was set to 0.037 \citep{Schl11}. The $R_V$ ratio for both {\tt ebv} components is set to 3.1.

The 12$^{\prime\prime}$ aperture of OM also takes in starlight in the optical band. Therefore, in our modeling, we incorporated a template model component for the stellar emission from the host galaxy's bulge, based on \citet{Kin96}, and allowed its normalization to be fitted. 

\begin{figure}
\centering
\includegraphics[width=0.53\textwidth]{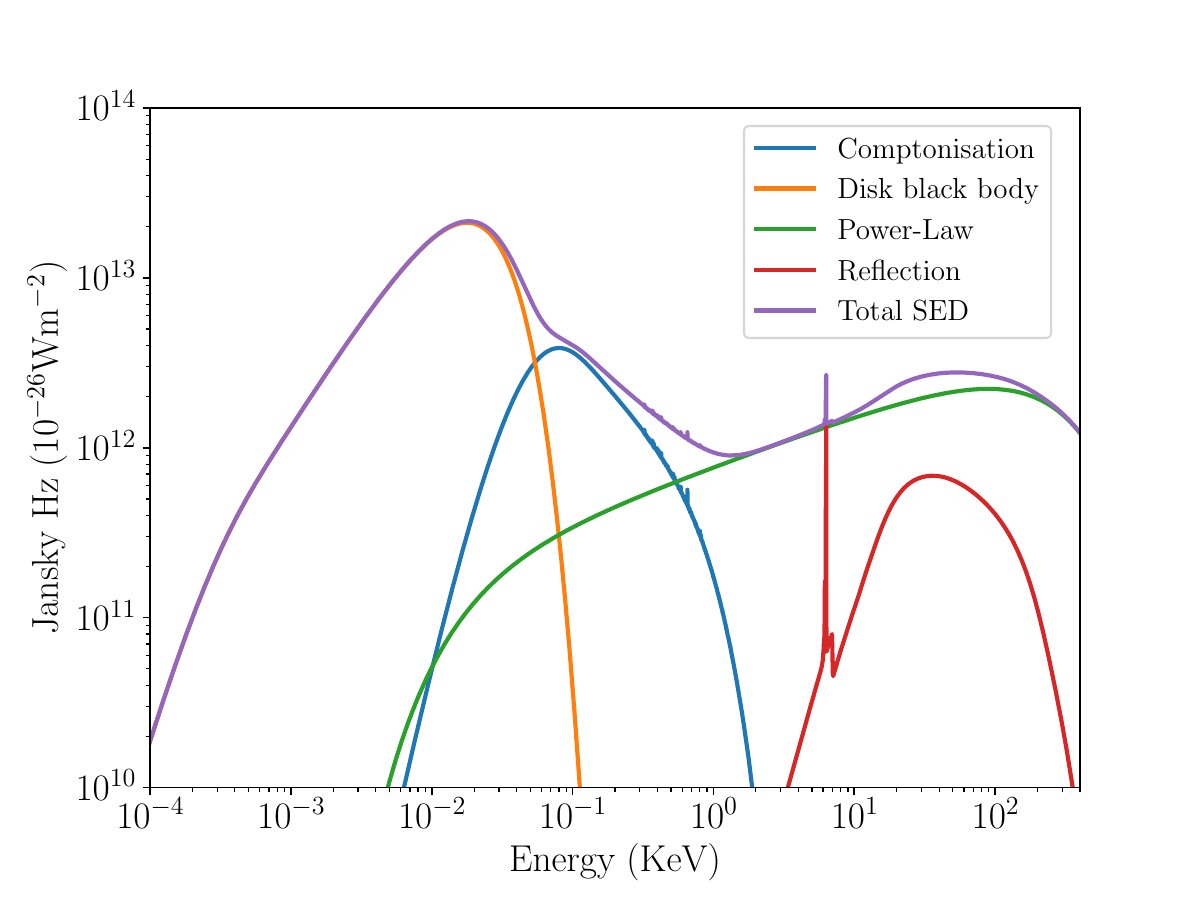}
\caption{The Spectral Energy Distribution (SED) of the fit includes multiple components: a disk black body, a power-law, a reflection model, and comptonization. The total SED is represented in purple.}
\vspace{-0.1in}
\label{fig:SED}
\end{figure}

\subsection{X-ray continuum modelling}

In addition to the optical/UV {\tt dbb} component (Sect. \ref{sect_optical}), the broadband continuum consists of three other components in the X-ray band: a power-law ({\tt pow}) component for modeling the hard X-ray continuum; an X-ray reflection component ({\tt refl}) for modeling the Fe K$\alpha$ line and the Compton hump at hard X-rays; and a component for the `soft X-ray excess' which we model with warm Comptonization using the {\tt comt} model in {\tt SPEX}. We keep the optical depth and the coronal temperature of the {\tt comt} component fixed to typical values \cite[e.g.][]{Petrucci2018} in order to avoid unnecessary free parameters which cannot be constrained.  

The \xmm and \chandra data were obtained at different epochs and therefore the continuum shape might be different due to variability intrinsic to the source. To take this variability into account, we use the \sectors option in \spex. With this option, each dataset is allocated to a specific sector. In this way, the continuum parameters (which are variable over time) for each dataset are treated independently and the normalisation of the \pow component is allowed to vary freely, which gives a more accurate description of the continuum. The free parameters of our continuum model, listed in Table \ref{table_cont}, are the normalisation of the \pow and \comt components, the photon index ($\Gamma$) of the \pow and the scale of reflection in \refl. The SED is plotted in Figure \ref{fig:SED}. The SED model we are using aligns well with the EPIC-pn data. We note that two \etau 
components in \spex are used to create the
low-energy and high-energy exponential cut-off of the power-law component at reasonable energies. A high-energy cut-off has been applied at 300 keV and a low-energy cut-off at the Lyman edge (13.6 eV).

\begin{table*}
\caption{Best fit parameters of the spectral energy distribution.}  
\label{table_cont}    
\begin{center}
 \small\addtolength{\tabcolsep}{+2pt}
%\scalebox{0.8}{%
\begin{tabular}{c c c }     
\hline\hline            
Spectral component                    &       Parameter          & Value                \\
\hline   
 \multirow{2}{*}{\pow}                &      \multirow{2}{*}{ Normalisation  ($\rm 10^{44} ph/s/keV$)}      &      $(2.4\pm0.1) \times 10^{7}$ : XMM-Newton     \\
                                      &                                                                      &    ($3.2\pm0.1) \times 10^{7}$ : Chandra   \\
                                      &     Photon Index $\Gamma$                      &     $1.69^{+0.02}_{-0.01}$           \\
\hline                
\multirow{4}{*}{\comt}                &    Normalisation  ($\rm 10^{44} ph/s/keV$)    &   $(1.7\pm0.3)\times 10^{10}$      \\
                                      &       $\rm T_{seed} (eV)$                    &   15 (coupled to \dbb)       \\
                                      &        $\rm T_{e} (keV)$                   &    0.2   (fixed)                   \\
                                      &     Optical depth $\tau$                      &     15   (fixed)                  \\

\hline                
\multirow{2}{*}{\refl}                &    Incident power-law Norm. ($\rm 10^{44} ph/s/keV$)      &  $3.2\pm0.1 \times 10^{7}$       \\
                                      &   Incident power-law $\Gamma$      &                1.7 (coupled to \pow) \\
                                      &    Reflection scale s               &      $ 0.24^{+0.08}_{-0.04}    $               \\

\hline               
\multirow{2}{*}{\dbb}                 &     Normalisation  ($\rm 10^{22} \ m^{2}$)  &   $9 \pm 2$             \\
 							     &        T (eV)                          &     15 (fixed)                          \\
\hline
\multirow{1}{*}{\tt ebv}                 &     $E(B-V)$ &   $0.45 \pm 0.04$             \\
\hline 
\end{tabular}
\end{center}
\begin{flushleft}
\footnotesize{Note: The {\tt ebv} component refers to the internal reddening in NGC 6860. The normalization of the \pow component has two distinct values, corresponding to the XMM-Newton and Chandra datasets respectively. When only one value is given it reflects both epochs.   }
\end{flushleft}
\end{table*}

\subsection{Photoionisation and spectral modelling of the ionised outflows}

The X-ray spectrum of NGC 6860 exhibits a series of absorption lines belonging to warm absorbers. We fit the RGS and HETGS spectra simultaneously using photoionisation modelling. For the RGS, we fit the spectral range of 7-35 \AA, while for the HETGS, we utilize both the High-Energy Grating (HEG) covering 1.6-14 \AA, and the Medium-Energy Grating (MEG) spanning 2.5-15 \AA. For the photoionisation modelling we choose the \pion model in \spex \citep{Mehdipour2016}. \pion is a self-consistent model that calculates the thermal and ionisation balance together with the spectrum of a plasma in photoionisation equilibrium. It uses the SED from the continuum model components set in \spex. During spectral fitting, as the continuum varies, the thermal and ionisation balance and the spectrum of the plasma are re-calculated at each stage. 
This implies that when employing realistic broad-band continuum components to analyze the data, the photoionisation is computed in accordance with the \pion model.

To adequately account for the absorption lines originating from various ionic species we need four $\pion$ components, with different ionisation parameters, $\xi$. The ionisation parameter is defined as $\rm \xi=L/ \nH \cdot r^{2}$, where L is the luminosity of the ionising source over the 1–1000 Ryd band (13.6 eV to 13.6 keV) in $\rm erg \cdot s^{-1}$, \nH \ the hydrogen density in $\rm cm^{-3}$, and $r$ the distance between the photoionised gas and the ionising source in cm (\citealt{Tarter1969}, \citealt{Krolik1981}). For the different \pion components we have as free parameters: the column density of the absorber, the ionisation parameters as well as the velocity shift and velocity broadening. The remaining \pion parameters have been kept at their default values, as specified in the user manual.\footnote{https://spex-xray.github.io/spex-help/models/pion.html}. In particular the density \nH of \pion is left at its default value as our modeling is not dependent on density. The pion components are added until all the observed absorption features are effectively modeled, and no further statistical enhancement in the fit is achievable. 

\begin{figure*}
\centering
%\begin{subfigure}
    \includegraphics[width=0.47\linewidth]{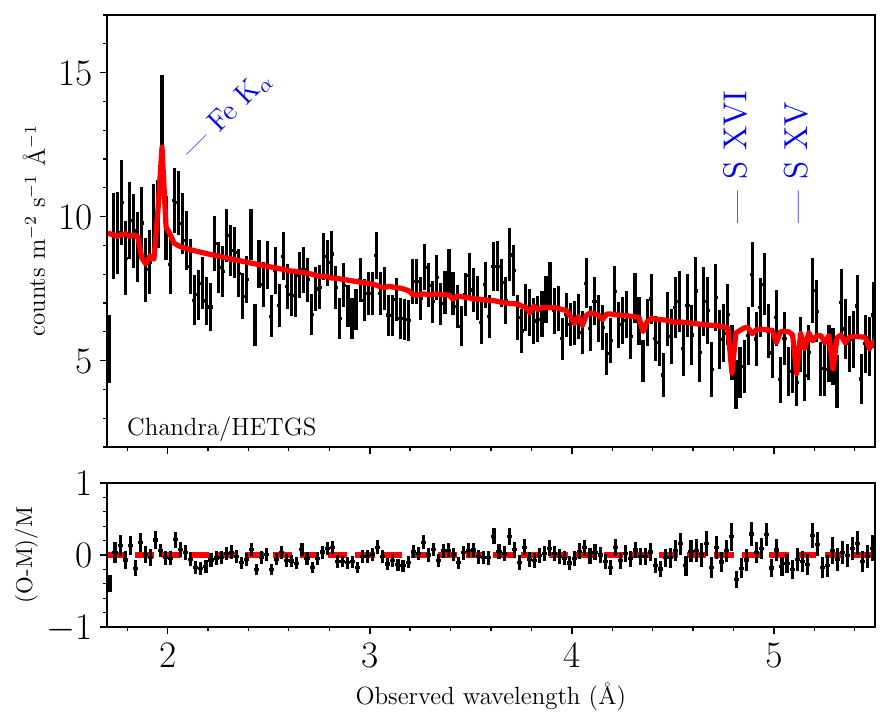}
    \includegraphics[width=0.47\linewidth]{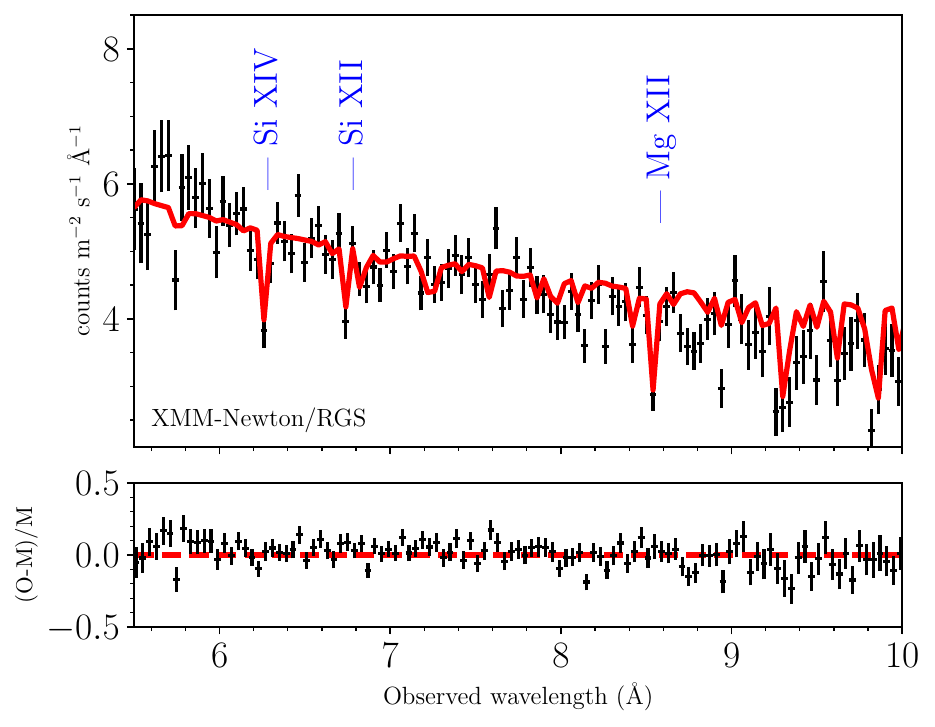}
    \includegraphics[width=0.47\linewidth]{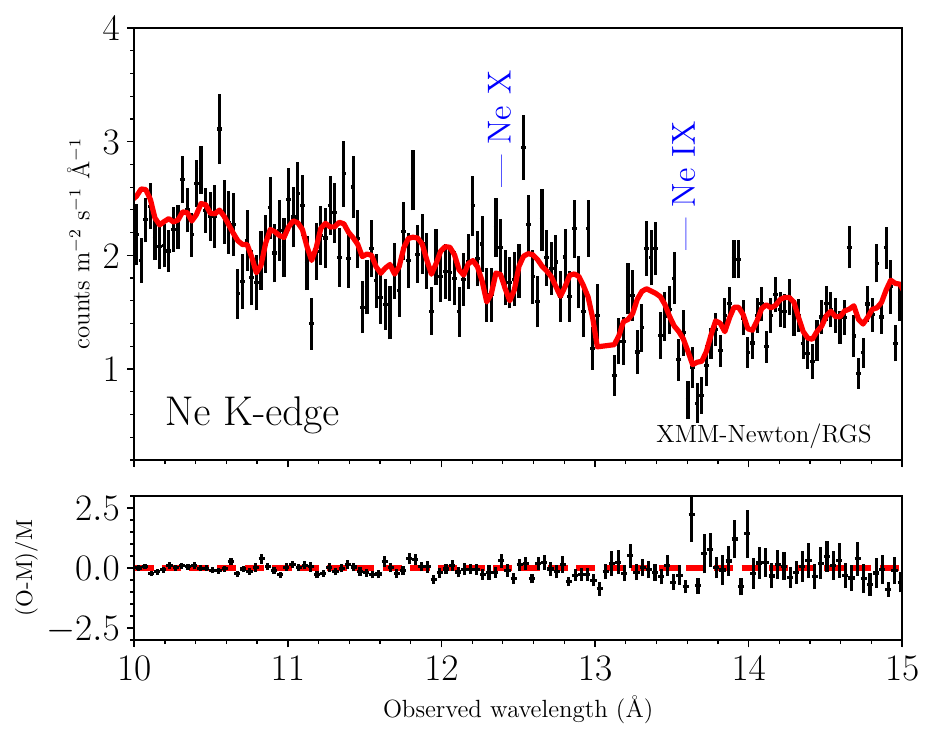}
    \includegraphics[width=0.47\linewidth]{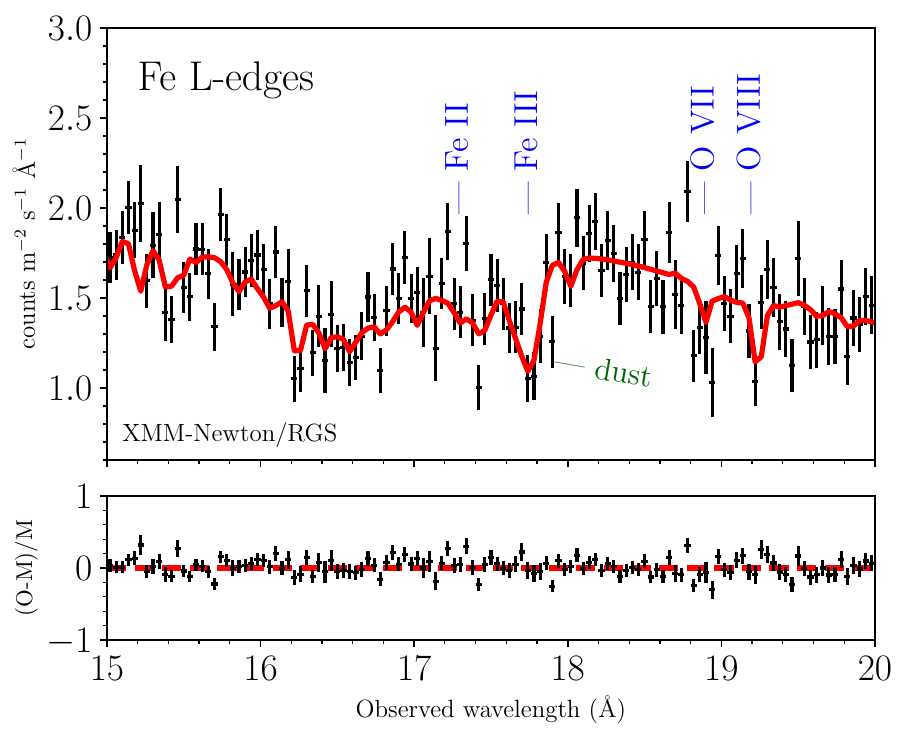}
    \includegraphics[width=0.47\linewidth]{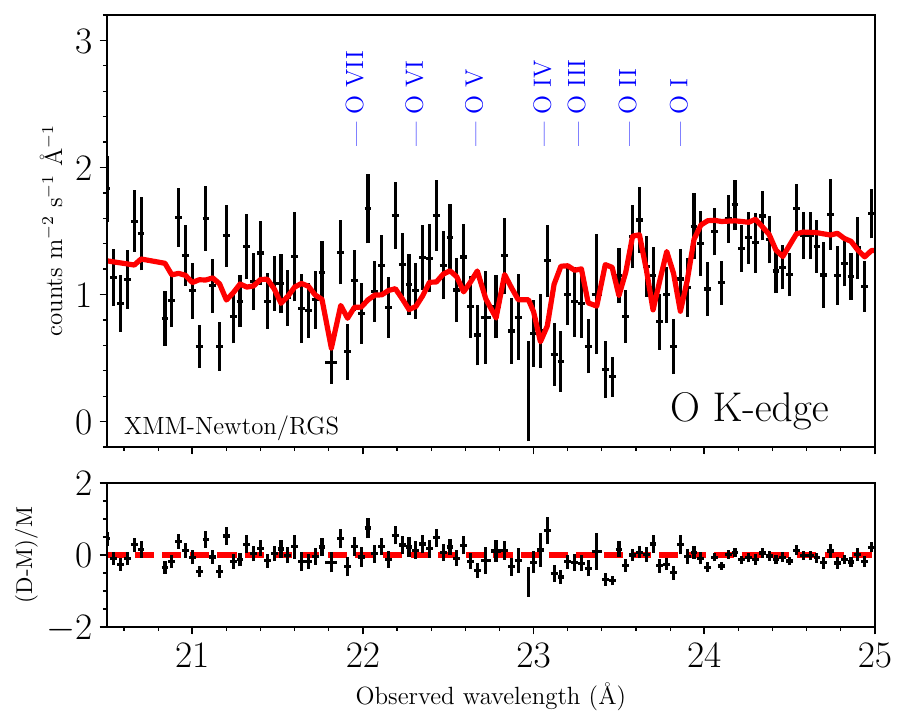}
%\end{subfigure}
\caption{High-resolution X-ray spectrum of NGC 6860. The panels show different spectral regions and absorption lines, including the $\rm Fe \ K\alpha$, and the Ne K-edge (10-15 \AA), Fe L-edges (15-20 \AA), O K-edge (21-25 \AA) spectral regions.}
\label{fig:spectra}
\end{figure*}

The highest ionisation component (Comp. A) of our fit, $\rm log\xi=3.4$, produces the absorption from highly ionised species such as \sxiv, \sxv, \fexix, \fexx, \fexxi, and improves our fit by $\Delta C_{stat}$=870. A second $\pion$ component (Comp. B) with lower ionisation parameter of $\rm log\xi=2.9$ fits the absorption lines of \ovii, \oviii, \neix, \nex and improves our fit by $\Delta C_{stat}$=530. A third $\pion$ \ component (Comp. C), $\rm log\xi=2.3$, is needed to improve the fit by $\Delta C_{stat}$=200. 
Finally we added a fourth weakly-ionized/neutral $\pion$ \ component (Comp. D) in the AGN. This component improves our fit by $\Delta C_{stat}$=1200 and contributes to the continuum absorption. Due to the lack of line detection, the velocity broadening ($\rm \sigma$) and velocity shift ($\rm v_{out}$) of this component are unconstrained. Therefore, we have fixed them at the default values, as they do not impact the fit. In our modeling the continuum first irradiates the highest-ionization \pion component and reaches the lowest-ionization component last. The best fit values of the \pion components are presented in Table \ref{pion_para}, and Figure \ref{fig:spectra} shows the best fit model on the data. Finally, we examined potential variability in the \pion components across different epochs using \xmm and \chandra observations. However, we found no conclusive evidence of variability. We also examined our spectra for the presence of emission lines. However, these features appear too weak for a meaningful analysis. Typically, the strongest emission lines would be \oviii $Ly\alpha$ and the \ovii forbidden line, but even these are barely detectable in our case. Our tests indicate that incorporating them into our modeling does not improve the fit, making it impractical to constrain their parameters.

A photoionized plasma can exhibit instability at certain ionization states. To determine whether our photoionization components are in pressure equilibrium, we create a stability curve (also known as an S-curve or cooling curve). The stability curve shows the change in the electron temperature as a function of the pressure form of the ionization parameter $\Xi$, introduced by  \citet{Krolik1981}, which is defined as the radiation pressure divided by the gas pressure.

We generate the S-curves by first constructing a grid of electron temperatures and ionization parameters $\xi$ based on our photoionization modeling and the SED in SPEX. We then compute the pressure, $\Xi$. The parameter $\Xi$ gives the ionization equilibrium and can be expressed as $\Xi =F/n_{H}\cdot c\cdot kT$, where $n_{\rm H}$ is the hydrogen density in $ \rm cm^{-3}$, $\it k$ is the Boltzmann constant, and  $\it T$ the gas temperature. Since $F=L/4\cdot \pi \cdot r^{2}$ and $\xi=L/(n_{H} \cdot r^{2})$, $\Xi$ can be written as

\begin{equation}
\Xi=\frac{L}{4 \pi  r^{2} \cdot n_{H} \cdot c \cdot kT}=\frac{\xi}{4 \pi \cdot c \cdot kT}=19222 \cdot \frac{\xi}{T}
\end{equation}

The stability curve is presented in Fig. \ref{fig:scurve}, where we also over-plot the location of our photoionisation components.

\subsection{Modelling of the dust X-ray absorption in NGC 6860}
\label{sec:AGNdust}

The host galaxy exhibits a notable reddening with a measured value of E(B-V) = 0.72 \citep{Lipari1993}, which is higher than the E(B-V) = 0.037 observed in our specific line of sight through the Milky Way. This substantial disparity highlights the presence of considerable dust extinction within NGC 6860. The RGS spectrum reveals intriguing residuals associated with the Fe LII and Fe LIII edges. Given that iron is highly depleted into dust (\citealt{Psaradaki2024}, and references therein), these features likely indicate the presence of iron-containing dust grains within NGC 6860.

The dust modifies the shape of the photoabsorption edge, giving a diagnostic for its detection in the X-ray spectrum. To model the dust X-ray absorption fine structures in the spectrum of NGC 6860, we employ the \amol \ model in \spex, which calculates the transmission of a dust component while treating the dust column density as a free parameter. Our methodology integrates the recently introduced dust extinction cross sections for the Fe L-edges (\citealt{Psaradaki2021}), derived from up-to-date laboratory data detailed in \citet{Psaradaki2021,Psaradaki2022}. These cross sections are computed using anomalous diffraction theory \cite[ADT,][]{van_de_Hulst} for the Fe L-edges, and Mie theory (\citealt{Mie}) for the O K-edge. A Mathis-Rumpl-Nordsieck dust size distribution \cite[MRN,][]{mathis} has been assumed for the computation of these models, which follows a power-law function, $dn/da\propto a^{-3.5}$, with grain sizes ranging from a minimum cutoff of $\rm 0.005 \ \mu m$ to a maximum cutoff of $\rm0.25 \ \mu m$. 

We include a composite dust mineralogy in our fitting procedure, consisting of metallic iron and pyroxene silicate ($\rm Mg_{0.75}Fe_{0.25}SiO_{3}$). This particular combination of dust components, identified as the most suitable compounds to describe the Fe L and O K edges according to \cite{Psaradaki2022}, is applied in our analysis. The inclusion of this model improves our fit by $\Delta C_{\text{stat}}$ =120. The parameters of the best fit dust component are listed in Table \ref{tab:paramsISM}. It is important to note that this model specifically accounts for dust in the vicinity of the AGN, as a reddening component has been taken into consideration. 
%{\color{red} can we test for other compounds? or rule out compounds?}

\subsection{Modelling the Milky Way ISM}

To account for neutral Galactic absorption, we utilize the $\hot$ model from \spex\ (\citealt{dePlaa}; \citealt{Steenbrugge2005}). This model calculates the ionization balance for a given temperature and set of abundances and then determines all ionic column densities scaled to the specified total hydrogen column density. At low temperatures (approximately 0.001 eV or 10 K), the $\hot$ model simulates a neutral gas in collisional ionization equilibrium. The fitted parameters are the hydrogen column density along the line of sight and the temperature ($kT$, where $k$ is the Boltzmann constant). This approach provides a physical fit of the ISM, allowing us to distinguish between different temperature phases along the line of sight. The depletion of silicon, magnesium and iron are constrained to be at least 90\% (\citealt{Zeegers2019}, \citealt{Rogantini2019}, \citealt{Psaradaki2022}), while the oxygen depletion is at 20\% (\citealt{Psaradaki2020}, \citealt{Psaradaki2022}).

For the neutral ISM component, we fix the hydrogen column density to the value of $\rm 3.1 \times 10^{20} \ cm^{-2}$ from \cite{Willingale2013}. The Galactic ISM also contains absorption lines of ionized species, particularly around the O K-edge (\citealt{Juett2004}; \citealt{Costantini2012}; \citealt{Pinto2013}; \citealt{Gatuzz2016}; \citealt{Psaradaki2020} and references therein), such as \oii\ and \oiii. Therefore, we add an additional $\hot$ component at a higher temperature to account for absorption lines such as \oii, \oiii, \oiv, \feiii, \feiv, \neiii, and \neiv. The results are presented in Table \ref{tab:paramsISM}.
Finally, we add an \amol\ model to account for Galactic Milky Way dust extinction, assuming a dust composition of a mixture of pyroxene and metallic iron, as mentioned in Section \ref{sec:AGNdust}. We find a total dust column density of $\rm <7.74 \times 10^{16}\ cm^{-2}$. The inclusion of these ISM components improves our fit by $\Delta C_{\text{stat}}$ = 80.

\begin{table}
\caption{Best fit parameters of the photoionisation component, \pion.}  
%\renewcommand{\arraystretch}{1}
%\begin{center}
 %\small\addtolength{\tabcolsep}{+2pt}
\scalebox{0.9}{%
\begin{tabular}{c c c c }     
\hline\hline            
Spectral component                     &       Parameter          & $\rm Value  $  & $\Delta C_{stat}$   \\
\hline   
 \multirow{4}{*}{\pion \#1}           &      \NH   &      $\rm 9.8^{+2.7}_{-5.7}$       &   \multirow{4}{*}{870} \\
                                      &     log$\xi$                       &  $\rm 3.43^{+0.16}_{-0.61}$                        \\
                                      &     $\rm v_{out}$                    &    $-188^{+51}_{-65} $                   \\
                                      &       $\sigma$           &   $70\pm28$                       \\
\hline     
\multirow{4}{*}{\pion \#2}            &     \NH                   &   $\rm 10.1\pm1.7$        &   \multirow{4}{*}{530}              \\
                                      &     log$\xi$                                           &   $2.9\pm0.1$                       \\
                                      &     $\rm v_{out}$                      &    $-280\pm31   $                  \\
                                      &   $\sigma$                      &    $54\pm14$                      \\
\hline                
\multirow{4}{*}{\pion \#3}            &        \NH                   &    $\rm 1.14\pm0.15$    &   \multirow{4}{*}{200}                        \\
                                      &       log$\xi$                                     &     $2.31\pm0.06$                     \\
                                      &       $\rm v_{out}$                   &    $-215^{+54}_{-63}$                      \\
                                      &    $\sigma $                      &  $46^{+23}_{-26}$                        \\
\hline               
\multirow{4}{*}{\pion \#4}            &        \NH                 &      $\rm 1.3\pm0.1$     &   \multirow{4}{*}{1200}                \\
                                      &      log$\xi$                                        &  $-5.7\pm0.7$                       \\
                                      &     $\rm v_{out}$                    &      -50 \ (fixed)                    \\
                                      &    $\sigma$                   &     100 \ (fixed)                    \\
\hline         
\end{tabular}}
%\end{center}
\label{pion_para}
\begin{flushleft}
\footnotesize{Note: The \NH is given in units of $\rm 10^{21} cm^{-2}$, log$\xi$ in $\rm ergs/ s\cdot cm$, $\rm v_{out}$   in $\rm km \cdot s^{-1}$, and $\sigma$ in $\rm km \cdot s^{-1}$.  }
\end{flushleft}
\end{table}

\begin{table}
\caption{Best-fit parameters for the \hot and \amol components, displaying the gas and dust column densities in the interstellar medium of the Milky Way (MW), as well as the dust column density near NGC 6860. The notation 'dust1' refers to metallic iron, while 'dust2' to amorphous pyroxene composition ($\rm Mg_{0.75}Fe_{0.25}SiO_{3}$).}  
%\renewcommand{\arraystretch}{1}
%\begin{center}
% \small\addtolength{\tabcolsep}{+2pt}
\scalebox{0.9}{%
\begin{tabular}{c c c }     
\hline\hline            
Spectral component                     &       Parameter          & Value                \\
\hline   
 \multirow{2}{*}{\hot \#1 MW}         &      \NH ($\rm 10^{20} cm^{-2}$)   &  $3.1$ (fixed)                       \\
                                      &        T (keV)              &    $\rm 10^{-6} \ (fixed)$                      \\
\hline                
\multirow{2}{*}{\hot \#2 MW}           &      \NH ($\rm 10^{20} cm^{-2}$)     &      $7.9^{+3.0}_{-0.3}$                  \\
                                      &     T (keV)                      &   $<0.01$                       \\
\hline                
\multirow{2}{*}{\amol \ MW}                &    $\rm N_{dust1}$ ($\rm 10^{16} cm^{-2}$)      & $<1.2$           \\
                                      &         $\rm N_{dust2}$ ($\rm 10^{16} cm^{-2}$)       &  $<6.54$       \\
\hline               
\multirow{2}{*}{\amol \ NGC6860}                  &   $\rm N_{dust1}$  ($\rm 10^{16} cm^{-2}$)    &    $2.3\pm1.3$        \\
 							    &          $\rm N_{dust2}$  ($\rm 10^{16} cm^{-2}$)            &  $10.1\pm0.5$        \\
\hline               
\end{tabular}}
%\end{center}
\label{tab:paramsISM}
\end{table}

% ------------------------------ Plots : spectra ----------------------------------- %

\begin{figure}
%\centering
%\hspace{-0.35cm}
\includegraphics[width=0.5\textwidth]{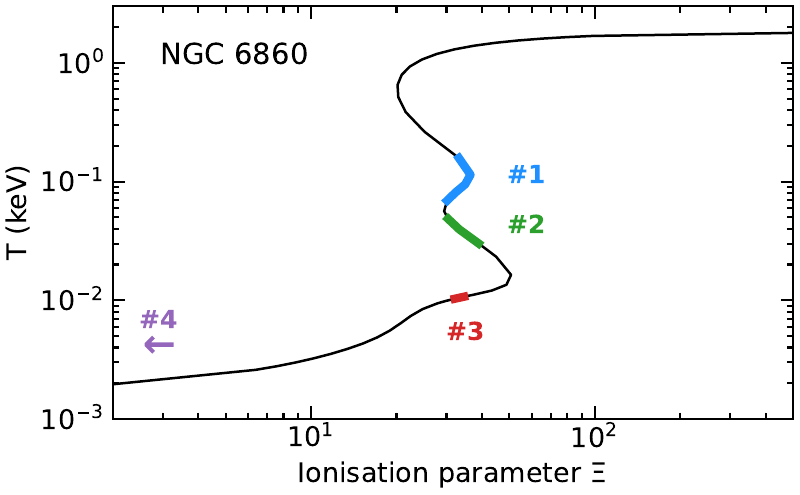}
\caption{The thermal stability curve (S-curve) of NGC 6860 computed with the {\tt pion} model. The location of the measured outflow components in NGC 6860, including their uncertainties, is highlighted as bands on the S-curve.}
\vspace{-0.1in}
\label{fig:scurve}
\end{figure}

\section{Discussion}\label{discussion}

\subsection{The ionised outflows in NGC 6860}\label{discussion:outflows}

From our modeling of the RGS and HETG spectra we find that the ionized outflows (or warm absorbers) in NGC 6860 consist of four components. There is a large difference between the ionization of components 1--3 and component 4, which is very lowly ionized. Assuming a constant velocity outflow with a volume filling factor ${f = 1}$, an upper limit for the distance ($r$) of the absorber from the ionising source can be obtained. This is done by substituting ${N_{\rm H} = n_{\rm H}\,r\,f}$ into the definition of the ionization parameter $\xi = L / n_{\rm H}\, r^2$, in order to eliminate the unknown density $n_{\rm H}$. Therefore, we get $r < L\, f / N_{\rm H}\, \xi$. By using the best-fit parameters of Table \ref{pion_para}, and $L = 1.7 \times 10^{44}$ erg~s$^{-1}$ from our SED modeling, we obtain ${r < 2}$~pc (\pion \#1), ${< 7}$~pc (\pion 2), and ${< 237}$~pc (\pion 3). For \pion 4, because $\xi$ is too low, the $r$ limit becomes practically unconstrained.

While the precise location of warm-absorber components are not known, the above distance constraints are consistent with typical locations near the torus (\#1 and \#2), the narrow line region (NLR, \# 3), and the extended NLR and the host galaxy (\# 4). The spatially-resolved optical spectroscopy by \citet{Benn06} finds that the NLR in NGC~6860 is extended out to 1.47 kpc. The \pion outflow component may have originated as thermal winds from the dusty AGN torus (e.g. \citealt{Krol01}). For the lowest-ionization component (\pion 4), the ionization and velocity of this gas cannot be well constrained. Therefore, it may have either originated as a torus wind, or alternatively this component is distant gas in the ISM of the host galaxy that is ionized by the AGN radiation. Warm absorbers by the host galaxy have been also found by \cite{DiGesu2017}. Interestingly, Fig. \ref{fig:scurve} shows that the highest ionization components (\pion \# 1 and 2) reside on the unstable parts of the S-curve. On the other hand the lower ionization components (\pion \# 3 and 4) are thermally stable. Furthermore, the \pion components \#1,2,3 have similar \ $\Xi$ values -- this would imply that these components are in pressure equilibrium. On the other hand, the gas described by \pion component \# 4, which has an extremely lower $\Xi$, appears to be in its distinct state and is likely separated from the other components. In the following we discuss the possibilities of dust presence in these warm-absorber components.

\subsection{The reddening in NGC 6860}

We find that the reddening in the host galaxy is $E(B-V) =0.45\pm0.04$. 
Using our reddening value, we can derive the hydrogen column density with the equation: $\rm N_{H} (cm^{-2}) = b \times E(B-V) (mag)$ where the b value ranges from 5.5 to 6.9 \citep{Drainebook, Predehl1995, Bohlin1978, Gorenstein1975, Guver2009}. This implies that the neutral hydrogen column density is approximately $2.5 \times 10^{21} \ \text{cm}^{-2}$ to $3.1 \times 10^{21} \ \text{cm}^{-2}$.

Apart from continuum modeling, another method to estimate the reddening is via measurement of the observed Balmer decrement (H$\alpha$/H$\beta$ flux line ratio). The extinction causes the observed Balmer decrement ($D_{\rm obs}$) to appear larger than intrinsic theoretical predictions ($D_{\rm int} \approx 2.85$, \citealt{Oster06}). Using the reported H$\alpha$ and H$\beta$ flux values for NGC 6860 from \citet{Malk17}, the observed Balmer decrement is 5.5. This  can be converted into reddening according to ${E(B-V) = a\, \log\, (D_{\rm obs} / D_{\rm int})}$, where $a$ depends on the choice of the extinction curve. For the SMC extinction curve \cite[$a = 1.475$, see][]{Mehdipour2018}, the inferred reddening would be 0.42. This matches the $E(B-V)$ that we measure from continuum modeling ($0.45\pm 0.04$, Table \ref{table_cont}).

Furthermore, the optical and near-infrared spectroscopic study of \citet{Lipari1993} reports that the reddening in the host galaxy of NGC 6860 is $E(B-V) = 0.72$. Using this larger value, the neutral hydrogen column density is approximately $4 \times 10^{21} \ \text{cm}^{-2}$ to $5 \times 10^{21} \ \text{cm}^{-2}$, depending on the b value. In our analysis, we determined the reddening by fitting the continuum using OM data. Discrepancies in $E(B-V)$ may also arise from various factors, such as the extinction curve applied, the energy range used for continuum modeling, and the type of continuum model that is used. Nonetheless, our results are consistent with the Balmer decrement method, described above.

The above $E(B-V)$ values all imply significant X-ray absorption in NGC 6860, in addition to the absorption of the Milky Way. However, from our X-ray modelling we do not require neutral absorption by the host galaxy (\hot spectral component). Furthermore, the neutral hydrogen column density that we obtained from our X-ray modeling with the \hot model is $3.03 \times 10^{20} \ \text{cm}^{-2}$, which is consistent with the Milky Way value of \citet{Willingale2013}: $\rm 3.58 \times 10^{20} \ \text{cm}^{-2}$. Therefore, the absorbing gas responsible for the internal reddening is likely the photoionised warm-absorber in NGC 6860, that we have modeled with the \pion. Interestingly, the column density that we measure with \pion is consistent with implied column density from the reddening.

\subsection{The nature and origin of dust}

A key indicator of dust X-ray absorption is that, although atomic and ionic absorption lines can be accurately fitted without considering dust, the photoelectric edges in the X-ray spectrum are not well-fitted. This occurs because dust affects the shape of the absorption edges. Therefore, additional absorption is needed to fit the K edge of O, and the LII, LIII edges of Fe.

To account for dust absorption in our fit, we included the \amol model in \spex, incorporating two different minerals: metallic iron and amorphous pyroxene silicate ($\rm Mg_{0.75}Fe_{0.25}SiO_{3}$). The dust is likely either embedded in the outflow, or present in the ISM of the host galaxy, or both. In our photoionisation modelling, we included four \pion components. The first two components, pion \#1 and \#2, are highly ionised with $\rm log\,\xi=3.43$ and $\rm log\,\xi=2.9$, therefore we do not expect dust to be embedded in these components.

The lowest-ionization \pion component (\#4) likely originates in the host galaxy of NGC 6860. Similar to our Milky Way, it is expected that the host galaxy contains dust, contributing to the observed absorption features. Unlike Milky Way, gas in the host galaxy of the AGN is ionized by the central AGN. This component is distinguished from our own Galaxy's dust absorption because it is redshifted, while our Milky Way's dust absorption component is not.
The lowest-ionization \pion component (\#4) from NGC 6860 exhibits a column density comparable to the values typically observed in the Milky Way \cite[e.g.][]{Psaradaki2022}. This similarity in column density further supports the presence of dust within the host galaxy.

The intermediate-ionization component (\pion \#3) originates from gas with an ionization parameter of $\rm log\xi=2.31$. This value places it in the thermally stable segment of the S-curve (Fig. \ref{fig:scurve}). Given its location, at a distance of less than 237 parsecs from the central source, this component is situated in the NLR (Sect.\ref{discussion:outflows}). The question is whether the warm absorber, corresponding to this intermediate-ionization component (\pion \#3), contains dust. Previous studies indicate that dust particles can survive destruction mechanisms such as sublimation and thermal sputtering under the conditions present in certain warm absorbers \cite[e.g.][]{Reynolds1997, Mehdipour2012}. Using the definition from \cite{Barvainis1987}, the sublimation radius (the minimum distance from the central source where grains can survive) is given by:

\begin{equation}
R_{\rm sub}=1.3 \ L^{0.5}_{\rm uv,46}\  T^{-2.8}_{1500} \ {\rm pc}
\end{equation}
%($7.6 \times 10^{43}$ erg s$^{-1}$)
where $\rm L_{uv,46}$ is the UV luminosity of the central source in units of $\rm 10^{46} \ erg \ s^{-1}$ (calculated from our SED), and $\rm T_{1500}$ is the grain sublimation temperature in units of 1500 K. Using the luminosity value in the UV band from our SED modeling, and the temperature of 1500K, we obtain a sublimation radius of $ R_{\rm sub}=0.01$~pc. So as long as the dust resides in a plasma which is farther than 0.01 pc from the source, it is not destroyed by sublimation.

Another mechanism that can destroy dust grains is sputtering, where high-energy collisions with hot ions cause atoms or molecules to be ejected from the surface of the dust grain, which gradually disintegrates the grain over time. Following \cite{Burke1974}, the sputtering threshold, which is the grain
lifetime against destruction by sputtering is

\begin{equation}
t_{\rm sput} = 6.25 \times 10^{11} \ (YnT^{0.5}_{4})^{-1} \ {\rm s}
\end{equation}
where $Y$ is the sputtering yield (the number of atoms ejected per incident particle) at $T_{4}$, the incident particle energy in units of $\rm 10^{4} \ K$, and $n$ is the incident particle density. According to our photoionisation modeling results (Fig. \ref{fig:scurve}), \pion \#3 has a temperature of 0.01 keV ($10^{5}$ K). The $n$ value can be obtained using the definition of the ionisation parameter ($\xi = L / n\, r^2$), the $r$ value from Sect. \ref{discussion:outflows}, and the sputtering yield from the formula $Y=0.01 \cdot (T_{4}/10^{5})$. We therefore estimate the sputtering timescale to be $t_{\rm sput}<1.3\cdot10^{13}$~s. 

We can then calculate how far the outflowing gas would travel to reach the sputtering timescale: ${r \sim t_{\rm sput} \cdot \rm v_{out}} $, where $\rm v_{out}$ is the outflow velocity of \pion \#3 (Table \ref{pion_para}). This assumes a simple steady flow. We thus find a distance of 83.6 pc. Therefore, if \pion 3 is located at $\rm r<83.6$~pc, it would survive sputtering. Since our derived distance for \pion \#3 is $r < 237$~pc (see Sect. \ref{discussion:outflows}), this means it is still feasible for dust to survive sputtering in this component.

\citet{Laha2016} explore the dynamics of warm absorbers and their impact on the host galaxy's environment. Assuming the observed outflow velocity of the warm absorber corresponds to the radial escape velocity ($\rm v_{out}$) at its distance from the central engine, the minimum launch radius ($\rm r_{min}$) can be estimated using the virial relationship:

\begin{equation} r_{min}=\frac{GM}{\rm v_{out}^{2}} \end{equation}

where the black hole mass $M$ for NGC~6860 is ${4.0 \times 10^{7}}$~$M_{\odot}$ \citep{Gonzalez2012}. Applying this formula, we find $\rm r_{min} = 4\ pc$. Under this assumption, the pion component 3 is located within the range $4 < r < 237$, where dust grains could potentially survive sputtering.

\subsection{Comparison with other AGN with dusty outflows}

Significant efforts have been made in the literature to investigate dust in active galactic nuclei (AGN). 
\citet{Lee2001} studied the dusty warm absorber in MCG-6-30-15 using Chandra/HETGS data. The presence of dust in the warm absorber of MCG-6-30-15 was previously identified by \citet{Reynolds1997}. They discovered that the warm absorber in this source has at least two zones: an inner warm absorber at distances characteristic of the broad line region (BLR) and an outer warm absorber at distances characteristic of the torus. The authors noted that dust cannot survive in the inner warm absorber due to the intense radiation field and must therefore reside in the outer absorber.
In NGC 6860, we identified at least four components. Two of these are consistent with typical locations near the torus, one component belongs to the narrow line region, and the fourth component is in the extended NLR and the host galaxy. Additionally, the dust grains in NGC 6860 can survive destruction (sublimation and sputtering) in the intermediate ionization warm absorber if it is located at a distance \( r < 83.6 \) pc. The dust is primarily located in the NLR and the torus.

\citet{Lee2001} found that the dust column density from their modeling aligns with the column density obtained from reddening studies, providing direct X-ray evidence for dust within a warm absorber. They suggest that the most plausible explanation for the detection of the Fe I L feature is absorption by dust embedded in partially ionized material. Similarly, in NGC 6860 there is significant absorption by dust in the Fe L. Dusty warm absorbers have also been proposed for other AGN, for example such as those discussed by \cite{Brandt1996}, \cite{Komossa1997, Komossa2000} and \cite{Lee2013}. 

\citet{Mehdipour2012} studied the warm absorber in the Seyfert 1.8 galaxy ESO 113-G010. Based on continuum modelling, dust-to-gas ratio arguments, and the measured properties of the ionized outflows, they concluded that dust is embedded in the AGN wind. This provides observational evidence that AGN winds likely transport dust into the ISM. Similar to NGC 6860, the authors found that dust grains can survive the sublimation and sputtering processes in the lower-ionised phase of the warm absorber and thus can cause the observed optical-UV reddening.

\citet{Mehdipour2018} conducted a multi-wavelength spectroscopic study to investigate the origin and nature of dust in the reddened quasar IC 4329A. There are two dust components in this object: an ISM dust lane in the host galaxy (seen in optical HST images), and a nuclear one (see in through X-ray absorption and reddening). The authors find that the nuclear dust is associated to dusty outflows from the AGN torus. The nuclear dust grains from the AGN torus and its wind in IC 4329A are likely larger than those in the Milky Way's ISM. They are in a porous composite form, containing amorphous silicate with iron and oxygen. Similarly, in NGC 6860, we find that the dust is primarily located in the NLR and the torus. However, additionally, we also see a dust component in X-rays that is located in the host galaxy (\pion \#4).

The recent literature suggests that a major portion of the mid-infrared (MIR) emission in AGN originates from an extended polar component, indicating the presence of polar dust \cite[][and references therein]{Asmus2019}. This polar dust is located at similar scales to the narrow-line region (NLR) of the AGN and likely forms a hollow, cone-like structure surrounding the NLR. \cite{Asmus2019} studied nine nearby obscured Seyfert-type AGN using deep, subarcsecond-resolution MIR imaging and found that polar dust may be a ubiquitous feature of AGN structure. Therefore, if polar dust resides on scales comparable to the NLR, our results may imply that the observed dust in NGC 6860 is outflowing polar dust. 

Finally, our study allows us to place  limits on the distance of the absorbers and investigate the presence and location of dust grains in this AGN. Previously, NGC 6860 was studied by \cite{Winter2010}, who analyzed Suzaku and XMM-Newton observations and identified a two-component warm ionized absorber in the soft X-ray spectrum. In our analysis, we identify three warm photoionised absorbers. This is likely due to differences in the coverage of high-resolution spectra (HETGS + RGS in this work), as well differences in modeling, which also impact the derived parameters.

\subsection{The physical origin of the ionised outflows}
One of the longstanding enigmas in AGN research is the nature and origin of ionised outflows (warm absorbers). Some studies have linked them to the presence of dust grains in AGN. \cite{Mizumoto2019} suggested that dust-driven winds may play a significant role in accelerating warm absorbers. Later on, \cite{Ogawa2021} investigated the ionization structure of the radiation-driven fountain model in a low-mass AGN \citep{Wada2016}. In this model, outflows are primarily driven by radiation pressure on dust and the thermal energy of gas resulting from X-ray heating, naturally forming a geometrically thick, torus-like structure. Their results suggest that additional components, originating closer to the SMBH than the torus—such as line-driven winds from the accretion disk—are likely crucial contributors to the formation of warm absorbers in AGN.  Our results on NGC 6860 support this scenario, as we find that the dust in this AGN likely originates from the NLR and is present in the warm absorber.

The importance of radiation pressure on dusty gas as a source of AGN feedback has also been discussed in the literature \cite[e.g.][]{Ishibashi2016, He}. Recently, \cite{Ishibashi2024} studied broad absorption line (BAL) outflows which are commonly detected in AGN but their driving mechanism is poorly understood. They concluded that dust-driven BAL outflows may provide a significant contribution to AGN feedback on galactic scales. In NGC 6860, we observe that the dust likely originates from the NLR and possibly the host galaxy, suggesting that the ionized dusty outflows could serve as a mechanism driving feedback within the host galaxy.  

\section{Conclusions}\label{conclusions}
In this study, we analyzed the high-resolution X-ray spectra of the active galactic nucleus NGC 6860 using high-resolution X-ray spectroscopy with XMM-Newton and Chandra. By applying photoionisation modeling, we determined the properties of the ionized outflows in NGC 6860, and investigated the nature and location of dust in this AGN. Our main findings are the following:

\begin{itemize}
    \item The RGS and HETG spectra reveal that the ionized outflows (or warm absorbers) in NGC 6860 consist of four distinct components. There is a significant disparity in ionization levels between components 1-3 and component 4, with the latter being very lowly ionized.  The first two components are highly ionized, with \(\log \xi = 3.43\) and \(\log \xi = 2.9\), respectively. Our thermal stability analysis showed that these components are not thermally stable. The third component is at an intermediate level, with \(\log \xi = 2.31\) and it belongs to a stable part of the S-curve. The fourth component is lowly ionized and is likely originating from the host galaxy.
    \item While the precise location of the warm absorbers is not known, we estimated an upper limit for the location (distance from the central black hole) of these components. Components 1 and 2 are located at $r<2$~pc, and $r<7$~pc respectively and are consistent with typical locations near the torus. The third component is at $r<237$~pc, and it is located in the narrow line region. The fourth component is located in the extended NLR and the host galaxy. 
    \item From our continuum modeling we measure an internal reddening of $E(B-V) =0.45\pm0.04$ in NGC 6860. The derived $E(B-V)$ value implies significant X-ray absorption in addition to that of the Milky Way. This absorbing gas is the photoionized outflows in the AGN. We find that the column density of this gas matches the column density expected from the reddening.
    \item We find that the dust is possibly located
in the NLR and the torus. Additionally, there is a potential dust component in the host galaxy. The dust grains in NGC 6860 can survive destruction mechanisms such as sublimation and sputtering in the intermediate-ionization warm absorber. If this component is located at $ r<83.6$~pc, it could survive sputtering.

\end{itemize}

%\begin{acknowledgments}

We would like to thank the referee for the suggestions that helped to improve the clarity of this paper. IP would like to thank Erin Kara for insightful discussions, and Jelle de Plaa for help with SPEX. This work is supported by NASA grant 80NSSC23K0490 for our XMM-Newton observation. This research has made use of data obtained from the Chandra Data Archive, and software provided by the Chandra X-ray Center (CXC). SRON is supported financially by NWO, the Netherlands Organization for Scientific Research. Support for this work was also provided by NASA through the Smithsonian
Astrophysical Observatory (SAO) contract SV3-73016 to MIT for Support
of the Chandra X-Ray Center (CXC) and Science Instruments. CXC is
operated by SAO for and on behalf of NASA under contract NAS8-03060.

This paper employs a list of Chandra datasets, obtained by the Chandra X-ray Observatory, contained in ~\dataset[DOI: https://doi.org/10.25574/cdc.327]{https://doi.org/10.25574/cdc.327}.

%\url{https://doi.org/10.25574/cdc.327}.

%\end{acknowledgments}

\bibliography{reference}{}
\bibliographystyle{aasjournal}

\end{document}